\documentclass{llncs}

\usepackage{amssymb,amsmath}
\usepackage{graphicx}
\usepackage{epic}
\usepackage{textcomp}
\usepackage{setspace}
\usepackage[usenames,dvipsnames]{color}

\begin{document}
\pagestyle{empty}
%\mainmatter

\title{Recommender System for Online Dating Service}
\titlerunning{Recommender System for Online Dating Service}

\author{Luk\'a\v s Bro\v zovsk\'y\inst{1}
\and V\'aclav Pet\v r\'i\v cek\inst{1}}
\authorrunning{Luk\'a\v s Bro\v zovsk\' y et al.}

\institute{KSI MFF UK\\
Malostransk\'e n\'am. 25,\\
Prague 1, Czech Republic\\
\email{lbrozovsky@centrum.cz, petricek@acm.org}}

\maketitle

\begin{abstract}
%at least 70 and at most 150 words.
%	Background
\hspace{1em}Users of online dating sites are facing information overload that requires them to manually construct queries and browse huge amount of 
matching user profiles. This becomes even more problematic for multimedia profiles.
%	Purpose
Although matchmaking is frequently cited as a typical application for recommender systems, there is a surprising lack of work published in this area.
%	Method & Results
In this paper we describe a recommender system we implemented and perform a quantitative comparison of two collaborative filtering (CF) and two global 
algorithms. Results show that collaborative filtering recommenders significantly outperform global algorithms that are currently used by 
dating sites. A blind experiment with real users also confirmed that users prefer CF based recommendations to global popularity recommendations.
%	Conclusion
Recommender systems show a great potential for online dating where they could improve the value of the service to users and improve 
monetization of the service.
\end{abstract}

%================================
\section{Introduction}
%================================
% setting
Users of online dating sites are facing information overload. A typical dating site requires users to fill in lengthy questionnaires and matching is 
based on attributes selected. A typical search in the database either returns no match in the case when the query is too specific or, for 
general queries, returns a huge amount of profiles. Ranking of the results is difficult and usually requires users to assign weight to each of their 
requirement. Such approach is limited especially because different users express the same thing in different words. In the end users end up browsing 
large amount of irrelevant results. Query suggestion and query rewriting has 
been used in information retrieval to alleviate this problem but it becomes even more problematic for multimedia profiles.

Offline dating services have a very long history. With the advent of the web some of these companies started to put up web sites and build their 
online presence. Many new startups appeared but generally they mimicked the processes of offline dating agencies including questionnaires and 
matching algorithms.\footnote{\url{http://www.date.com}, \url{http://www.match.com}, \url{http://www.perfectmatch.com}, etc.} In the last few years, a 
specific subgroup of online dating service web applications has emerged on the Internet. Their main idea is to let users store personal profiles and 
then browse and rate other users' profiles. The registered user is typically presented with random profiles (or with random profiles preselected with a 
simple condition - for example only men of certain age) and rates the profiles on a given numerical scale.\footnote{\url{http://hotornot.com}, 
\url{http://libimseti.cz}, \url{http://chceteme.volny.cz}} Online dating sites generally allow users to post and browse profiles for free 
and require payment for contact details. From the business model point of view the dating agencies should have high interest in improving the quality of recommendations they provide.
Recommender systems~\cite{resnick97recommender} have been used in many areas. They have been popularized especially by recommenders at Amazon, 
Netflix, Movielens and others. Even though matchmaking is often cited as a potential area for recommender systems application there has been a 
surprising lack of work published in this area.

% purpose
In this paper we describe a recommender system for online dating agency, benchmark algorithms on real world dataset and perform a study with real 
users to evaluate the quality of recommendations.  Presented improvement in online dating and matchmaking has many benefits for users of the service 
as well as for the owners. These benefits include higher user satisfaction and loyalty, and also a better monetization of the service.

% This paper is organized as follows. We first describe the algorithms we implemented (Section~\ref{}). We then describe the dataset(Section~\ref{}) 
% that we used for algorithm evaluation (Section~\ref{}). User experiment setup and results are then presented in Section~\ref{}. Finally Section~\ref{} 
% presents a summary of our contributions  and future work.

%================================
\section{Related Work}
%================================
%5.1 RS
%5.1.1 Resnick
%5.1.2 popular applications
%5.1.2.1 amazon
%5.1.2.2 Netflix
%5.1.3 most successful RS are CF based
%5.1.4 history of RS
%5.1.4.1 tapestry
%5.1.4.2 Ringo
%5.1.4.3 Grouplens Movielens
%5.1.4.4 launch
%RS history
Recommender systems~\cite{resnick97recommender} are a popular and successful way of tackling the information overload. Recommender systems have been popularized by applications such as Amazon \cite{linden03amazon} or Netflix recommenders\footnote{\url{http://amazon.com}, \url{http://netflix.com}}.  The most widely used recommender systems are based on collaborative filtering algorithms. One of the first collaborative filtering systems was Tapestry~\cite{goldberg92using}. Other notable CF systems include jester~\cite{goldberg01eigentaste}, Ringo~\cite{upendra94social}, Movielens and Launch.com.

%5.1.5 CF
%5.1.5.1 memory based CF
%5.1.5.2 model based
%5.1.5.2.1 SVM
%5.1.5.2.2 MMMF
%5.1.5.2.3 PCA
%5.1.5.2.4 Bayes nets
%5.1.5.2.5 item item
%5.1.5.2.6 personality diagnosis
There are two classes of CF algorithms - memory based (e.g. user-user) and model based. Memory based systems require the whole rating matrix stored in memory. Typical example of a memory based algorithm is user-user k-nearest neighbor algorithm. Model based algorithms include methods based on rating matrix decomposition: for example SVD~\cite{sarwar02incremental}, MMMF~\cite{rennie05fast}, and other methods like item-item~\cite{sarwar01itembased}, Bayes-networks~\cite{breese98empirical}, personality diagnosis~\cite{pennock00collaborative}.
Collaborative filtering research has been addressing many areas including prediction accuracy, scalability, cold start problem, robustness and recommendation quality.
Breese et al.~\cite{breese98empirical} and \cite{herlocker04evaluating} contain empirical comparisons of different types of algorithms in terms of prediction MAE and review of results in this area.
 Deshpande et al.~\cite{deshpande04item} performed a comparison of topN item based recommender systems on several datasets including movies, credit card transactions, and others using metrics such as hit rate.
In addition to these cross-validation experiments several studies with real users were performed.
Ziegler et al~\cite{ziegler05improving} showed how topic diversification in the recommended set increases book recommendation quality as perceived by users. Recommendation explanation role has been recognized and/or used in \cite{cosley02referee,sinha01comparing,kautz97referral}.  Cosley et al.~\cite{cosley02referee} compared several recommender systems in CiteSeer digital library.
Despite the amount of previous work none of it addressed dating in particular and none of the datasets used represented dating preferences. 
%5.1.6.2 explanation
%5.1.6.2.1 referee
%5.1.6.2.2 referral web
%5.1.6.2.3 friends v system

Even though dating is often cited as a potential application for collaborative filtering based recommender systems and some dating sites claim to use proprietary collaborative techniques\footnote{http://www.match.com} there is a surprising lack of literature published on this problem.

%==============================
\section{Algorithms}
%==============================
%3.1 random
%3.1.1 predicts rating randomly
%3.1.1.1 The Random Algorithm is the simplest model-based recommender
%3.1.1.2  Its every prediction is an uniformly
%3.1.1.3 distributed random value within the ratings value scale. For fixed user a and item j it shall
%3.1.1.4 always predict the same random rating value pa,j . Keeping this constraint ensures that
%3.1.1.5 algorithm's behavior is invariable even
%3.1.1.6 when involved in the independent runs.
\subsubsection{Random Algorithm}
 is a simple algorithm -- every prediction is an uniformly distributed random value within the rating scale. For 
fixed user $a$ and profile $j$ it always predicts the same random rating $p_{a,j}$. This ensures that the predictions though random 
stay the same between runs.

\subsubsection{Mean Algorithm}
is sometimes referred to as the Item Average Algorithm or the POP algorithm~\cite{breese98empirical}. The prediction $p_{a,j}$ is calculated as the mean value of all the non-zero ratings 
for profile $j$ ($\overline{r_{\centerdot{},j}}$). This algorithm is not personalized and completely ignores the information entered by the active 
user $a$. Random and Mean algorithms serve as our comparison baselines.

\subsubsection{User-User Algorithm}\label{labUserUser}~\cite{herlocker99algorithmic} is one of the best known collaborative filtering algorithms. 
When predicting ratings for active user $a$, the user database is first searched for users with similar ratings vectors to the 
user $a$ -- `neighbors'. 
Ratings of the $k$ most similar neighbors who rated item $i$ are then used to calculate the 
prediction for the active user $a$ and profile $j$:

\begin{equation}
\label{eqPrediction}
p_{a,j}=\overline{r_a}+\xi \sum_{i=1}^{k}w(a,n_i)(r_{n_i,j}-\overline{r_{n_i}})
\end{equation}

where $n_i$ is the $i$-th nearest neighbor, $\xi$ is a normalizing factor, and $\overline{r_{n_i}}$ is the mean rating of the user $n_i$.  
As similarity we used the well known 
Pearson's correlation coefficient~\cite{resnick94grouplens}:

\begin{equation}
\label{eqPearson}
w(a,j)=\frac{\sum _i(r_{a,i}-\overline{r_a})(r_{j,i}-\overline{r_j})}{\sqrt{\sum _i(r_{a,i}-\overline{r_a})^2\sum _i(r_{j,i}-\overline{r_j})^2}}
\end{equation}

where the summations over $i$ are over the profiles which both users $a$ and $j$ have rated. 

Our User-User algorithm implementation is parametrized by by two parameters i) $MinO$ -- minimum number of common ratings between users necessary to calculate user-user similarity and ii) $MaxN$ -- maximum number of user neighbors to be used during the computation. We refer to User-User algorithm with parameters $MinO=5$, $MaxN=50$ as ``User-User (5,50)''.

\paragraph{Item-Item algorithm}~\cite{sarwar01itembased} uses a different view on the ratings matrix. Instead of utilizing the similarity 
between rows of the matrix as User-User does, it computes similarities between profiles.\footnote{Please note that in the dating service environment, 
both users and items could be seen as representing users.} We used adjusted Pearson correlation as item-item similarity:

\begin{equation}
\label{eqPearsonAdj}
w_{adj}(j,l)=\frac{\sum _i(r_{i,j}-\overline{r_i})(r_{i,l}-\overline{r_i})}{\sqrt{\sum _i(r_{i,j}-\overline{r_i})^2\sum _i(r_{i,l}-
\overline{r_i})^2}}
\end{equation}

where the summations over $i$ are over the users who rated both profiles $j$ and $l$.
Each pair of common ratings of two profiles comes from a different user with 
The adjusted version of the Pearson correlation subtracts each
user's mean -- otherwise the similarity computation would suffer from the fact that
different rating scale.
 When making prediction $p_{a,j}$ for the active user $a$,
the ratings of the active user $a$ for the $k$ most similar neighbors to profile $j$ are used:

\begin{equation}
\label{eqPredictionII}
p_{a,j}=\overline{r_{\centerdot{},j}}+\xi \sum_{i=1}^{k}\tilde{w}(j,n_i)(r_{a,n_i}-\overline{r_{\centerdot{},n_i}})
\end{equation}

where $n_i$ is the $i$-th most similar profile to profile $j$ that user $a$ rated,
$\xi$ is a normalizing factor, and $\overline{r_{\centerdot{},u}}$ 
is the mean rating of profile $u$.

Similar to User-User algorithm, our implementation of Item-Item algorithm is also parametrized by two parameters i) $MinO$ -- minimum number of common ratings between profiles necessary to calculate item-item similarity and ii) $MaxN$ -- maximum number of item neighbors to be used during the computation.
We refer to the Item-Item algorithm with parameters $MinO=5$, $MaxN=50$ simply as 
`Item-Item (5,50)'.

%=====================================
\section{Libimseti Dataset}
%=====================================
The dataset we used consists of data from a real online dating service -- Libimseti. A snapshot 
of the rating matrix from July 2005\footnote{\url{http://www.ksi.ms.mff.cuni.cz/\~{}petricek/data/}} was used in this study. In the dataset users appear in two different roles: i) as active providers of ratings for 
photographs of others -- we will denote them as `users' in this situation and ii) as objects on photographs when they are rated by others -- we will 
refer to them in this situation as `profiles'.

\begin{figure}[h!]
\centering
\input{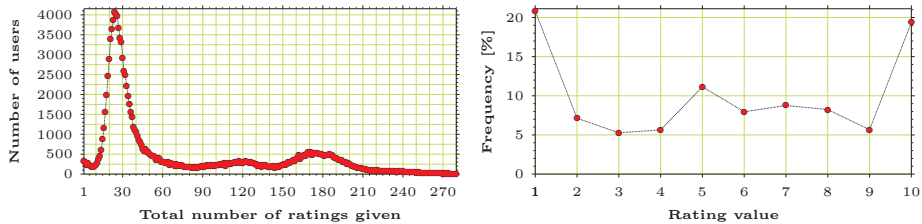}
\caption{Distribution of number of ratings. On the left is the distribution of number of ratings provided by single user. On the right then the distribution of the rating value.} \label{fig:dist-ratings}
\end{figure}

\begin{figure}[h!]
\centering
\input{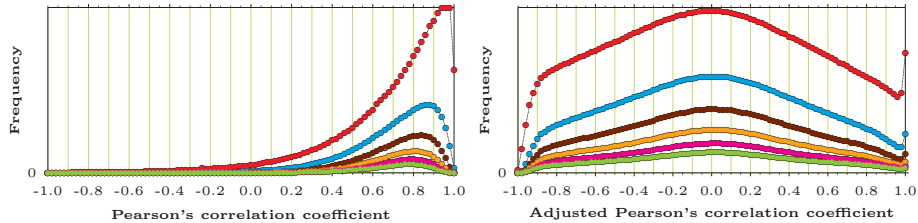}
\caption{Similarities distributions. User-User similarity and Item-Item similarity distribution. Each graph displays 6 distributions for different minimum overlap parameter (from top to bottom $MinO$ = 5,10,15,20,25,30).} \label{fig:dist-similarities}
\end{figure}

Overall the dataset contains 194,439 users, who provided 11,767,448 ratings. The sparsity of the matrix is 0.03\%. Table~\ref{tabDataSetsOverview} 
compares our dataset to the Movielens~\cite{herlocker99algorithmic} and Jester~\cite{goldberg01eigentaste} datasets that are well known in collaborative filtering 
research community. We can see that Libimseti is much larger but also sparser than the other two datasets. The distribution of number of ratings 
provided by a single user and distribution of the rating value are shown in Figure~\ref{fig:dist-ratings}. The similarity distributions for ratings 
matrix rows and columns are shown in Figure~\ref{fig:dist-similarities}.

\begin{table}[h!]
\centering
\caption{Data sets overview. An overview of the data set's basic characteristics. An \textit{active user} is user who have rated another profile at least once. The last three rows contain statistics of ratings normalized to $0-1$ scale.} \label{tabDataSetsOverview}
\begin{tabular}{|l|c|c|c|c|}
\hline
 & \textbf{MovieLens} & \textbf{Jester} & \textbf{LibimSeTi} \\
\hline
total users & 9,746 & 73,521 & 194,439 \\
\hline
users with ratings& 6,040 & 73,421 & 128,394 \\
\hline
items with ratings & 3,706 & 100 & 120,393 \\
\hline
ratings & \textbf{1,000,209} & \textbf{4,136,360} & \textbf{11,767,448} \\
\hline
density & 10.5302 \textperthousand & 0.7652 \textperthousand & 0.3113 \textperthousand \\
\hline
max ratings from 1 user & 2,314 & 100 & 15,060 \\
\hline
max ratings for 1 profile\hspace{1em} & 3,428 & 73,413 & 69,365 \\
\hline
rating (mean/med/sd)& 0.64/0.75/0.27  & 0.54/0.57/0.26 & 0.50/0.44/0.36 \\
\hline
\end{tabular}
\end{table}

%============================================
\section{ColFi Recommender System}
\begin{figure}[h!]
\centering
\label{figColFiArchitectureDesignOverview}
\includegraphics[width=\textwidth]{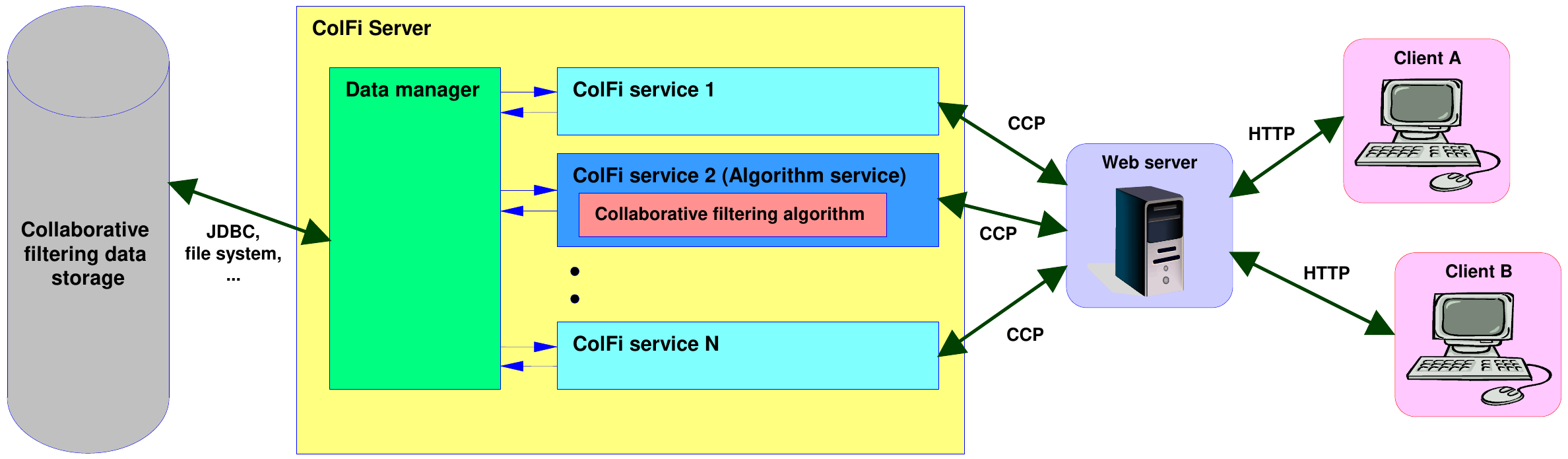}
\caption{ColFi architecture. The communication protocol between the data storage and the data manager is implementation dependent.}
\end{figure}

%============================================
We implemented a domain independent and freely available\footnote{http://colfi.wz.cz} recommender system that has been tested on four different datasets~\cite{colfi06}.
ColFi System architecture has been designed to be flexible yet simple enough so that developers can focus on collaborative filtering algorithms. Core 
of the system is a stateless TCP/IP server. It consists of a global data manager, a set of ColFi services and a communication module implementing CCP 
- ColFi Communication Protocol. The data manager serves as the main data provider for all the components in the system and ColFi services expose API 
to clients). Particular data manager and ColFi services implementations are provided as server plug-ins. 
Figure~\ref{figColFiArchitectureDesignOverview} summarizes ColFi architecture design.

System was implemented as a stateless TCP/IP client-server solution to achieve simplicity, platform independence, scalability and good performance. 
Each ColFi service instance has its own unique TCP/IP port number and is independent of all the other ColFi services. Multiple client connections are therefore supported and handled in parallel.  
We implemented the system in Java. The implementation allows easy addition of collaborative filtering algorithms. We use a lazy approach 
to similarity computations and store the similarities in a local cache object. MySQL server is used as a back end but the database is fully abstracted 
in the code. For more details on the implementation see~\cite{colfi06}.

ColFi Communication Protocol is similar to popular Java RMI. CCP is a very simple protocol and it supports only invocation of limited set of remote methods. That makes it very fast 
unlike SOAP
Webservices) and does not require special setup on the client side unlike Java RMI, EJB, JSP, or Servlets.

Data Manager is the heart of the ColFi system. It is a general interface that provides all the other ColFi components with data necessary for 
collaborative filtering. There is always exactly one data manager instance per server instance in a form of a plug-in for the ColFi server and the actual 
behavior can differ significantly for different implementations. The data manager can for example cache data in the memory or it can directly access 
database or special file system. In some other cases, read only implementation may be desired. Although both cached and direct access is implemented, 
only cached variants achieve reasonable performance. Direct access is generally usable only with small data sets.

ColFi Service is the only part of the system exposed to the outside world. Each service has its own TCP/IP server port number where it 
listens for incoming client requests. ColFi services are stateless, so each request is both atomic and isolated and should not affect any other 
requests. This allows simple parallel handling of multiple client connections on the server. The ColFi service is not limited to do collaborative 
filtering, it can also provide statistics for example.

%==============================
\section{Benchmarks}
%==============================
We compare the implemented algorithms on Libimseti dataset using three types of cross-validation. In each of the scenarios we provide algorithms with 
one subset of ratings for training and withhold other subset of ratings for testing the prediction accuracy.

%=================
\subsection{Setup}
%=================
We used three types of cross-validation: i) AllButOne ii) GivenRandomN and iii) production. Each validation uses NMAE as a metric of 
prediction quality:

\begin{align}
\label{eqNMAE}
\mathrm{NMAE} = \frac{\frac{1}{c}\sum_{k=1}^{c}|\tilde{p}^{k}_{ij}-r_{ij}|}{r_{max}-r_{min}}
\end{align}

It is possible that some algorithms fail to predict certain ratings due to lack of data. These cases are ignored and not included in NMAE 
computation\footnote{Our experiments showed that this does not affect the ordering of algorithms.} and only the total number of such 
unsuccessful predictions is reported.

\subsubsection{AllButOne Validation} is the most commonly used benchmark among many collaborative filtering 
papers~\cite{goldberg01eigentaste,breese98empirical}. It is the simplest protocol of the three. The idea behind this protocol is to select exactly one 
rating and hide it from the tested algorithm. Algorithm is then asked to predict that one hidden rating and the prediction error is the absolute 
difference between the hidden rating value and the predicted rating value. We reapeat this for \textit{every} single rating in the test data set.

\subsubsection{GivenRandomX Validation} is intended to examine the algorithm's performance with less data available from the active user. This 
situation is sometimes referred to as user cold-start and happens when a user is new to the system and did not provide enough ratings yet.

The validation works on subset of users who provided more than 100 ratings. We split the ratings into two: a training group T containing 99 random 
ratings from each user and a test set with remaining ratings. Group T is then used to generate 99 training sets $t_i$ for $i=1..99$ in such a way that 
$\bar t_i \bar = i$ and $\forall_i t_i \subset t_{i+1}$. These sets represent the ratings gradually provided by active user. The result of the 
GivenRandomX validation is a graph of the NMAE values as a function of training set size $i$.

\subsubsection{Production Validation} is designated to test the algorithm as a part of the ColFi System. The entire test data set is split into two 
disjunctive sets: the initial set and the simulation set. Both sets are about the same size. ColFi server is started with an algorithm being tested. 
Exactly $N_{max}$ clients (users) insert ratings from the simulation set of the test data set into the initial set through the ColFi server interface (this 
includes additions of new users to the systems as well). Ratings from the simulation part are selected at random. Prior to every single rating 
insertion on the server side, a prediction for the rating being inserted is computed. After every $K$-sized block of insertions, the NMAE of that 
set/block is calculated.  
The Production validation result is a graph of NMAE values based on the index of the 
$K$-sized block of insertions.

%===================
\subsection{Results}
%===================
Benchmarks were run on PC, 2x AMD Opteron\texttrademark{}~250 (2.4GHz, 1MB), 8~GB RAM, running Red Hat 
Linux~3.2.3-47. The variable parameters of the protocol were set as follows: $K$~=~10,000 and $N_{max}$~=~100 (hundred concurrent clients).

\subsubsection{AllButOne Validation Results} are summarized in Table~\ref{tabAllButOne}. First, an expected observation, that bigger neighborhood 
results in a better accuracy of the User-User algorithm. Further experiments did confirm that more than 50 neighbors do not significantly improve 
the performance. Item-Item algorithm with neighborhood size of 50 was used for the benchmarks.
Interesting observation is the small difference in NMAE values for Mean and CF algorithms. 
Overall User-User algorithm has best performance. Mean algorithm performs surprisingly well due to strong global component in user preferences.

\begin{table}[h!]
\centering
\caption{AllButOne results. Columns contain: i) NMAE of the algorithm prediction calculated in the data set's original scale ii) Number of all skipped rating predictions (due to insufficient data)} \label{tabAllButOne}
\begin{tabular}{|l|c|c|}
\hline
\textbf{algorithm} & \textbf{NMAE} & \textbf{skipped} \\
\hline
Random~~~ & ~39.72\%~ & ~0~ \\
\hline
Mean~~~ & ~15.69\%~ & ~24,785~ \\
\hline
User-User (5,10)~~~ & ~14.54\%~ & ~74,560~ \\
\hline
User-User (5,50)~~~ & ~13.85\%~ & ~74,560~ \\
\hline
User-User (10,10)~~~ & ~13.56\%~ & ~174,352~ \\
\hline
User-User (10,50)~~~ & ~13.34\%~ & ~174,352~ \\
\hline
Item-Item (10,50)~~~ & ~14.06\%~ & ~252,623~ \\
\hline
\end{tabular}
\end{table}

\subsubsection{GivenRandomX Validation Results} 
\begin{figure}[h!]
\centering
\includegraphics[width=\textwidth]{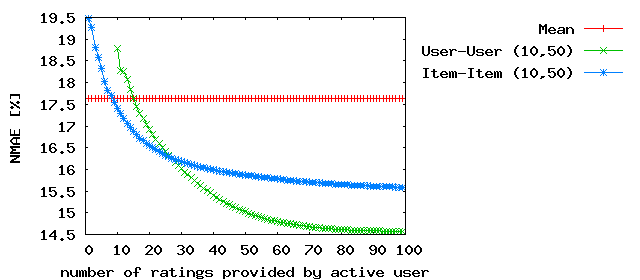}
\caption{GivenRandomX benchmark results. NMAE values based on the number of ratings given by a potential new user. The graph shows the behavior of 
the algorithms as they get more information about particular user. Data for the Random algorithm are not displayed as it has relatively high constant 
NMAE value of $38.27\%$.}
\label{figGivenRandomLibimSeTi}
\end{figure}

are presented as NMAE graph in the Figure~\ref{figGivenRandomLibimSeTi}. For readability the Random 
algorithm results are not displayed.

Algorithms show improved performance in terms of NMAE values for the first few ratings 
inserted. After the user enters about 30-40 ratings, algorithm's accuracies get stabilize (algorithm has enough information to derive the 
active user type). 
Overall, the User-User algorithm has generally lowest NMAE values.

\subsubsection{Production Validation Results} 
\begin{figure}[h!]
\centering
\includegraphics[width=\textwidth]{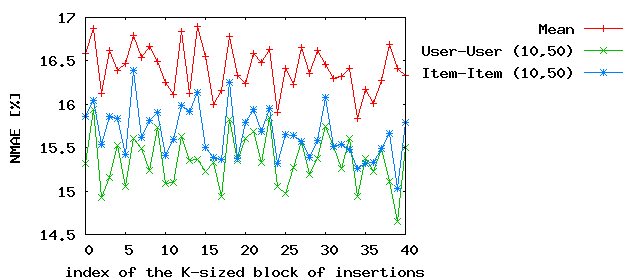}
\caption{Production benchmark results. NMAE values based on the index of $K$-sized set/block of insertions. Data for the Random algorithm are not 
displayed as it has relatively high mean NMAE value of $39.12\%$.}
\label{figProductionLibimSeTi}
\end{figure}

are presented in the Figure~\ref{figProductionLibimSeTi}. Results for the worst performing Random 
algorithm are again not displayed in the NMAE graphs to improve the legibility of the figure. 
 The performance ordering of algorithms corresponds to the AllButOne and GivenRandomN results.
During the course of this experiment we also verified that ColFi recommender can handle 
significant load at the level of real world application traffic.

%===================================
\section{User experiment}
%===================================
We conducted a web based experiment with human subjects. Each participant was asked to provide ratings for 150 random 
profiles. Based on the ratings provided we generated two recommendation lists of top 10 profiles.
We tested the Random, Mean, and User-User algorithms. User-User (10,50)\footnote{MinO=10, MaxN=50} was selected as the best performing collaborative algorithm, Mean represented the currently deployed functionality at Libimseti.cz and Random served as a baseline.
Recommendation lists contained only profiles that 
the active user did not rate yet. Each of the recommendation lists was generated using one of the ColFi algorithms. The two lists were then presented 
to the user who was asked to selected the list that he perceives as better recommendation. The order of the recommendation lists was randomized to 
avoid bias. Gender information was used to provide opposite sex recommendations.

Participants of the study were 111 users recruited mainly through the collab mailing 
list.\footnote{\url{http://www2.sims.berkeley.edu/resources/collab/}} From July to August 2006 we collected 14,057 ratings for 1,000 profiles (matrix sparsity 11.39\textperthousand).
Table~\ref{tabEmpirical} 
summarizes the outcomes of individual ``recommendation duels'' between all pairs of algorithms tested.
User-User algorithm was perceived by users as best although global mean algorithm has won a surprising fraction of duels.

\begin{table}[h!]
\centering
\caption{User study results -- outcomes of duels between the three algorithms tested. Each cell contains the percentage of duels where algorithm on the left won over algorithm on top. Cells where algorithm on left was percieved better more than 50\% of time are in bold.  \label{tabEmpirical}}
\begin{tabular}{|l|c|c|c|c|c|}
\hline 
& \textbf{~Random~} & \textbf{~Mean~} & \textbf{~User-User~}  \\
\hline
\textbf{Random~~} & - & 16.87\% & 12.50\% \\
\hline
\textbf{Mean~~} & \textbf{83.13\%} & - & 35.62\% \\
\hline
\textbf{User-User~} & \textbf{87.50\%} & \textbf{64.38\%} & - \\
\hline
\end{tabular}
\end{table}

The Random algorithm ``lost'' against both collaborative filtering algorithm and the Mean algorithm. On the other hand, the fractions won in the table for the 
Random algorithm -- 12.5\% and 16.87\% -- are quite high, which indicates that percieved profile quality is clustered due to self-selection bias (good looking people are more likely to put 
their photo online). The surprising success of the Mean algorithm -- 35.62\% wins against collaborative filtering algorithm -- suggests a strong universal preference that is shared by users -- in fact this is a prerequisitive for any beauty contest to be possible.

%================================
\section{Discussion}
%================================
We have shown that collaborative filtering based recommender systems can provide good recommendations to users of online dating services. We 
demonstrated this using cross-validation on a snapshot of a real world dating service. User-User and Item-Item algorithm outperformed global popularity in terms of 
prediction NMAE $3.08\%$ and $2.04\%$ respectively. We also verified that this difference in recommendation quality is noticable to users 
who preferred recommendations by User-User algorithm to global popularity recommendations.

A logical continuation of our work is a more complete evaluation of current state of the art recommender systems including model based collaborative 
filtering methods in the setting of dating services. These methods could also help to address scalability and performance issues. Domain specific improvements may be possible.

User interface may introduce bias in the sense that users instead of providing their personal preference try to guess the global preference. This 
reduces the usefulness of ratings provided. It remains an open issue how to best design an interface which motivates users to provide sufficient 
amount of truthful ratings. Application of index structures to speed up nearest neighbor search is an interesting research direction.

Recommendations can be further improved by hybrid algorithms. These algorithms are combining the collaborative filtering approach with content 
information. Another problem specific to dating is that ``A likes B'' does not imply ``B likes A''. Therefore each user should be probably presented 
with recommendations of such users, who are also interested in him/her. There is a need for reciprocal matching algorithms.

%========================
\section{Acknowledgments}
%========================
We would like to thank our three anonymous reviewers for their ideas and suggestions.
We would also like to thank Old\v rich Neuberger for providing the anonymized Libimseti dataset and to Tom\'a\v s Skopal for 
his valuable comments. This work was partially supported by the Ministry of Education of the Czech Republic grant MSM0021620838.

\begin{spacing}{0.9}
\bibliographystyle{plain}
\bibliography{colfi-znalosti-2007}

\begin{thebibliography}{10}

\bibitem{breese98empirical}
John~S. Breese, David Heckerman, and Carl Kadie.
\newblock Empirical analysis of predictive algorithms for collaborative
  filtering.
\newblock pages 43--52, 1998.

\bibitem{colfi06}
Luk\'{a}\v{s} Bro\v{z}ovsk\'{y}.
\newblock Recommender system for a dating service.
\newblock Master's thesis, KSI, MFF UK, Prague, Czech Republic, 2006.

\bibitem{cosley02referee}
Dan Cosley, Steve Lawrence, and David~M. Pennock.
\newblock {REFEREE}: An open framework for practical testing of recommender
  systems using researchindex.
\newblock In {\em 28th International Conference on Very Large Databases, {VLDB}
  2002}, 2002.

\bibitem{deshpande04item}
Mukund Deshpande and George Karypis.
\newblock Item-based top-n recommendation algorithms.
\newblock {\em ACM Trans. Inf. Syst.}, 22(1):143--177, 2004.

\bibitem{goldberg92using}
David Goldberg, David Nichols, Brian~M. Oki, and Douglas Terry.
\newblock Using collaborative filtering to weave an information tapestry.
\newblock {\em Commun. ACM}, 35(12):61--70, 1992.

\bibitem{goldberg01eigentaste}
Ken Goldberg, Theresa Roeder, Dhruv Gupta, and Chris Perkins.
\newblock Eigentaste: {A} constant time collaborative filtering algorithm.
\newblock {\em Information Retrieval}, 4(2):133--151, 2001.

\bibitem{herlocker99algorithmic}
Jonathan~L. Herlocker, Joseph~A. Konstan, Al~Borchers, and John Riedl.
\newblock An algorithmic framework for performing collaborative filtering.
\newblock In {\em SIGIR '99: Proceedings of the 22nd annual international ACM
  SIGIR conference on Research and development in information retrieval}, pages
  230--237. ACM Press, 1999.

\bibitem{herlocker04evaluating}
Jonathan~L. Herlocker, Joseph~A. Konstan, Loren~G. Terveen, and John~T. Riedl.
\newblock Evaluating collaborative filtering recommender systems.
\newblock {\em ACM Trans. Inf. Syst.}, 22(1):5--53, 2004.

\bibitem{kautz97referral}
H~Kautz, B~Selman, and M~Shah.
\newblock Referral web: combining social networks and collaborative filtering.
\newblock {\em Communications of the ACM}, 1997.

\bibitem{linden03amazon}
G~Linden, B~Smith, and J~York.
\newblock Amazon.com recommendations: item-to-item collaborative filtering.
\newblock {\em Internet Computing, IEEE}, 2003.

\bibitem{pennock00collaborative}
David Pennock, Eric Horvitz, Steve Lawrence, and C.~Lee Giles.
\newblock Collaborative filtering by personality diagnosis: A hybrid memory-
  and model-based approach.
\newblock In {\em Proceedings of the 16th Conference on Uncertainty in
  Artificial Intelligence, {UAI} 2000}, pages 473--480, 2000.

\bibitem{rennie05fast}
JDM Rennie and N~Srebro.
\newblock Fast maximum margin matrix factorization for collaborative
  prediction.
\newblock {\em Proceedings of the 22nd International Conference on Machine
  Learning}, 2005.

\bibitem{resnick94grouplens}
P.~Resnick, N.~Iacovou, M.~Suchak, P.~Bergstorm, and J.~Riedl.
\newblock {GroupLens: An Open Architecture for Collaborative Filtering of
  Netnews}.
\newblock In {\em {Proceedings of {ACM} 1994 Conference on Computer Supported
  Cooperative Work}}, pages 175--186. ACM, 1994.

\bibitem{resnick97recommender}
Paul Resnick and Hal~R. Varian.
\newblock Recommender systems.
\newblock {\em Commun. ACM}, 40(3):56--58, 1997.

\bibitem{sarwar02incremental}
B~Sarwar, G~Karypis, J~Konstan, and J~Riedl.
\newblock Incremental singular value decomposition algorithms for highly
  scalable recommender systems.
\newblock {\em Fifth International Conference on Computer and Information
  Science}, 2002.

\bibitem{sarwar01itembased}
Badrul~M. Sarwar, George Karypis, Joseph~A. Konstan, and John Riedl.
\newblock Item-based collaborative filtering recommendation algorithms.
\newblock In {\em Proceedings of the tenth international conference on World
  Wide Web}, pages 285--295, 2001.

\bibitem{sinha01comparing}
Rashmi~R. Sinha and Kirsten Swearingen.
\newblock Comparing recommendations made by online systems and friends.
\newblock In {\em {DELOS} Workshop: Personalisation and Recommender Systems in
  Digital Libraries}, 2001.

\bibitem{upendra94social}
S.~Upendra.
\newblock Social information filtering for music recommendation, 1994.

\bibitem{ziegler05improving}
Cai-Nicolas Ziegler, Sean~M. McNee, Joseph~A. Konstan, and Georg Lausen.
\newblock Improving recommendation lists through topic diversification.
\newblock In {\em WWW '05: Proceedings of the 14th international conference on
  World Wide Web}, pages 22--32. ACM Press, 2005.

\end{thebibliography}
\end{spacing}
\end{document}